# Seamless Long Term Learning in Agile Teams for Sustainable Leadership

M. R. J. Qureshi and M. Kashif
Department of Computer Science,
COMSATS
Institute of Information Technology, Lahore.
rjamil@ciitlahore.edu.pk, sheikhmkashif@yahoo.com

*Abstract*—**Seamless and** continuous support for long term organizational learning needs is essential for long lasting progress of the organization. Agile process model provides an excellent opportunity to cater that specific problem and also helps in motivation, satisfaction, coordination, presentation and technical skills enhancement of agile teams.

This long term learning process makes organization to sustain their current successes and lead both organization and team members to successful and dynamic market leaders.

## I. INTRODUCTION

Computer software era is marked with significant and fast changes. If an organization has to progress as well as sustain in this challenging, evolving and competitive environment. It has to predict and focus on long term needs and trends.

Long term learning requirements can be foreseen from organizational future plans, maintenance contracts and new emerging technologies. On the other hand learning desires of the agile team members can be discovered during agile extensive meetings and team members can also be asked formally for there specific learning.

Agile is excellent in terms of time reduction, and focus on people makes it fit with in software industry environment.

Features of agile as small team and reduced time, makes possible to move people from one team to another for learning purposes on the basis of learning opportunities, long term organizational needs and learning desires of team members.

As team members are moving from team to team it will lead to coordination, knowledge sharing and satisfaction of team members. These advantages/enhancements will lead team and organization to the market leader.

To support this view, related work (Sec. II) was reviewed (Sec. III) and proposed solution is summarized at (Sec. IV) then a study (Sec. V) was conducted with goals; how to find long term organizational needs (Sec. VI.A), finding team members learning desires (Sec. VI.B), finding learning opportunities with in organization and assigning it to appropriate team members (Sec. VI.C). The role of this learning process in terms of motivation, coordination, communication, satisfaction, and presentation skills were also studied (Sec. VI.D) and it is concluded (Sec. VII) that results support our solution.

## II. RELATED WORK

Lynne Ralston [1], very effectively describes 10 values that a person must have to lead an agile team. However, this list in not exhaustive and encourages audience to define their own list. Research focuses on integrity, relations, scope and need, seamlessly foster learning, authority, people, encouragement; fair play with customer, in-time delivery and team building.

Agile principles [2] are applicable to both (IT based & non IT based organization) for operations and leadership. Continuous implementation of agile lead to the development of a pool of leaders who transitioned into many leadership positions. Authors have used the concept of meme (An idea that exists in the mind of people), created distinction of memetic material and find the fittest meme for further use. Following meme was introduced and sustained leadership culture. These are; short cycle releases for customer satisfaction, incremental improvement to deliver good enough product today, daily team meetings to effectively overcome communication issues and reducing the risk of technical problem by preliminary technical experiments to find solutions.

Scrum's nature of self organization [3] helps to improve employee skills. Improving skills along with tight schedule is difficult with traditional approaches for organizational management. However, scrum backlogs and daily meetings leads to improvement of employee skills. Skill improvement responsibility in scrum is given to employees, who take decisions according to their needs/ambitions. However role of manager is not fully delegated, he/she can intervene according to need.

Knowledge sharing [4] is at the heart of the software development. This increases employee skills and makes them motivated and satisfied in long run. It is even more important in agile projects due to its short nature. Scrum provides the ways for intra team knowledge sharing, while inter team knowledge sharing is basically being done by factory experience method (Structured approach for knowledge sharing). Authors have combined both structured (ES) and unstructured (MASE) approaches for knowledge sharing. EB combined with MASE provide process oriented knowledge sharing on one hand and support for informal learning regardless of teams location on the other hand. But this is time intensive task.

This paper [5] tries to look into the long term impact of agile methods on software development life cycle. Currently it is very difficult to assess the long term impact of agile. Proponents of the agile were of the view that it is rapid, cost effective and efficiently manage the change and quality related issues. It helps to build team's qualities and manage the project properly. Opponents viewed that agile methodologies are narrow focused. These do not focus on software architecture which makes maintenance difficult and also results in higher error rates that increases the cost. It is customer centric leading to coordination and communication over burden.

Effect of agile on large team's performance [6] and skills is studied by using "scrum of scrums" technique. Analysis focuses on; autonomy of developers in daily tasks, variety of tasks, significance of every member, daily feedback and completion of whole tasks. These factors lead to learning, motivation and job satisfaction, thus improving employee skills effortlessly.

Agile seamlessly lead [7] to learn through knowledge sharing. Four processes are proposed for knowledge sharing i.e. pair programming, the pin point introduction of pairs is necessary during programming. Co-location may be of open environment or bullpen type makes communication and knowledge sharing easy and instantaneous. Daily status meetings not only help to keep project on track but also sharing of knowledge. Documentation is reduced as much as possible, instead a knowledge sharing tool (Wiki web, share point portal etc) is used.

XP (type of agile) heavily supports learning and build/enhances the leadership qualities of team members [8]. Learning becomes easy by observing, discussing and sharing knowledge/experiences with each other. XP enhance planning, communication and presentation skills Pair programming is also useful for reliving the members from communication problems. As the team working collectively so it is the responsibility of whole team to make the project a success, this increases the sense of responsibility on each and every team member.

Agile have all basic characteristics [9] that makes it effective tool for motivation and leadership. These factors make the projects be delivered in time and cost effective. Agile provide a result driven team structure and collaborative environment. Where all members compete and share knowledge for the sake of common goal (Project Success). As the objectives for agile teams are high it forces every person to do its best to achieve these goals (Customer satisfaction, time and cost effective releases). All these factors enhances the team a leadership qualities.

As agile [10] lead team to be a market leader so it must be sound in socio-psychological factors. Agile provides the speed and ease culture making team nimble and efficient. It helps to control in group and out group biases. It makes team to plan for collective goal, make estimations on the basis of resources and constraints in order to make the flexible and personalized planning, which results in process improvement.

It is evident that a number of researchers are working on the agile and its effect on the leadership, but very little is available on the incorporation of the agile for satisfying long terms organizational needs.

### III. PROBLEM STATEMENT

Agile support of learning for leadership focus on short term goals and does not focus on long term needs of the organization.

### IV. PROPOSED SOLUTION

There is need to resolve the agile problems in agile way instead of introducing structured approaches into agile which will indeed hurt the spirit of agile.

The proposed solution is consists of three phases and effect of them on leadership qualities.

### A. Find out Long Term Learning Requirements

An organization must foresee and predict the future trends in order to take decisions for leadership. Project manager can predict it by utilizing existing project documentations such as maintenance plan, maintenance contracts and marketing plan. He may also coordinate with the marketing team to find out the predicted future projects and approximate time required for their maturity. Currently running projects and their maintenance requirements are also helpful to predict future learning requirements. History, project contracts and market trends are also helpful in this connection. Emerging technologies have a paradigm effect on the learning requirements so they must be taken in to consideration for predicting future learning requirements.

### B. Ascertain the Learning Desires and Find the Opportunities.

Agile provide an excellent opportunity of interaction among all stake holders. Learning desires/needs of the team members can be found during these interactions e.g. through regular scrum review meetings. The team members can also be asked formally to show their area of learning desire. On the basis of this a list of learning desires is formed. Opportunities for these desires may be found/ searched within organization in different agile teams.

### C. Matching (Decision & Training)

Till now we have found long term learning requirements of an organization along with learning desires of the team members and opportunities of the learning with in organization so there are three lists of findings.
   a. A list of predicted learning requirements.
   b. A list of learning desires of the team members.
   c. A list of opportunities of learning with in organization.

Those learning desires which match with future work requirements will result into narrow downing of the learning list. On this list a further intersection is applied with learning opportunities found in the organization. This will results into final list providing the information which team member have to be trained what (long term learning requirements) and where (learning opportunities with in organization).

In order to, not to hurt agile spirit and avoiding any disturbance in the organization it is proposed that only a small number of the employees may be in this process at any given time. This learning process will results into followings.

As people will move from one team to another for learning purpose it will results in to increase inter team interaction. As a result of increased futuristic learning team members will remains satisfied and motivated.

These learning requirements are predicted for future by focusing on long term industry needs it will results into long term and focused learning.

Team members are moving from one team to another for learning purposes therefore they will coordinate with each other; they will communicate and will be presenting their ideas with the different nature of the people. All these are qualities that a progressing organization must have in its members. Team members are leaving and entering in software organizations. When they leave other members have to fill his hole. If an organization is adopting learning for leadership it will be an easy task for their members to fill this hole easily. Therefore in this way learning for leadership provides a backup support for organization to handle such type of cases.

## V. VALIDATION OF THE PROPOSED SOLUTION

A questionnaire consisting of 34 close ended questions divided into 4 goals was used for data gathering on basis of likert scale. Questions were arranged according to their relevancy to defined goals. Once data is collected it is statistically analyzed and interpreted to find support to our hypothesis or vice versa.

## VI. FINDINGS

Following are the following findings as per defined goals.
### A. Cumulative Statistical Analysis of Goal 1. Find out Long Term Learning Requirements.

Long term learning requirements are the foundation of this whole learning for leadership process. Once they are predicted properly they help a lot to have a clear idea; in which direction an organization is moving. It is also of paramount importance from the aspect of sustaining the leadership role because proper futuristic learning will make an organization able to sustain its market leader status. The result of the analysis of the goal 1 is shown in table I.

TABLE I
CUMULATIVE STATISTICAL ANALYSIS OF GOAL 1

| Q. No | Str. Disagree | Disagree | Neutral | Agree | Str. Agree |
|---|---|---|---|---|---|
| 1 | 4 | 7 | 6 | 15 | 3 |
| 2 | 1 | 2 | 8 | 17 | 7 |
| 3 | 2 | 4 | 5 | 17 | 7 |
| 4 | - | 10 | 9 | 13 | 3 |
| 5 | - | 11 | 10 | 10 | 4 |
| 6 | 2 | 8 | 6 | 16 | 3 |
| 7 | - | 4 | 12 | 14 | 5 |
| 8 | - | 3 | 13 | 17 | 2 |
| 9 | - | 4 | 7 | 17 | 7 |
| 10 | 1 | 1 | 1 | 11 | 21 |
| Total | 10 | 54 | 77 | 147 | 62 |
| Avg. | 2.85 | 15.42 | 22 | 42 | 17.71 |

As it is clear from the cumulative descriptive analysis of goal 1 that 42% of the sample agreed that we can find the long term learning requirements with the proposed solution and 17.71 % strongly agreed to it.14.52% disagreed to it and 2.85% strongly disagreed to it while 22% remained neutral as shown below in "Fig. 1".

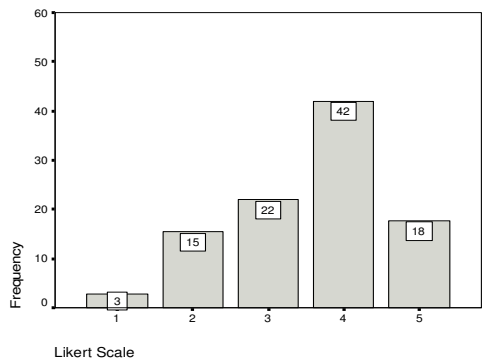

Figure 1. Graph showing the cumulative results of questionnaire for goal 1

*B. Cumulative Analysis of Goal 2. Ascertain the Learning Desires and Find the Opportunities.*

After finding the organization long term learning requirements, two more learning related issues are sorted in goal 2. We found the learning desires of the team members and learning opportunities with in organization. The cumulative result of survey for goal 2 is shown in table II.

TABLE II
CUMULATIVE ANALYSIS OF GOAL 2

| Q. No | Str. Disagree | Disagree | Neutral | Agree | Str. Agree |
|---|---|---|---|---|---|
| 1 | 1 | 1 | 8 | 16 | 9 |
| 2 |  | 2 | 3 | 18 | 12 |
| 3 |  | 2 | 2 | 21 | 10 |
| 4 | 1 | 3 | 7 | 14 | 10 |
| 5 |  | 2 | 3 | 18 | 12 |
| Total | 2 | 10 | 23 | 87 | 53 |
| Avg. | 1.14 | 5.71 | 13.14 | 49.71 | 30.28 |

Learning desires and opportunities can be found. 49.71% of the people were agreed to it and 30.28% strongly agreed to it.5.71% of the people disagreed and 1.14% strongly disagreed to it. While 13.14% remained neutral as shown in "fig. 2".

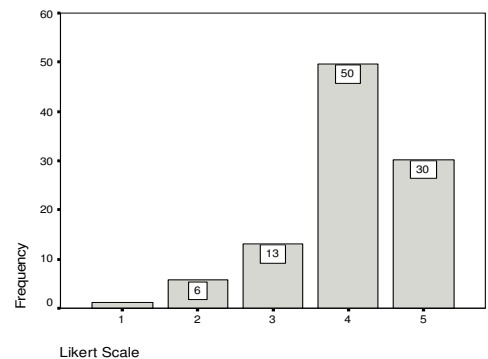

Figure 2. Graph showing the cumulative results of questionnaire for goal 2

*C. Cumulative Analysis of Goal 3. Matching (Decision & Training)*

This goal provides the answers of the questions that will be trained for leadership? Where will be trained? Why will be trained? The result of survey of this question is shown below in table III.

TABLE III
CUMULATIVE ANALYSIS OF GOAL 3

| Q. No | Str. Disagree | Disagree | Neutral | Agree | Str. Agree |
|---|---|---|---|---|---|
| 1 |  | 2 | 6 | 20 | 7 |
| 2 | 1 | 12 | 11 | 8 | 3 |
| 3 |  |  | 14 | 15 | 6 |
| 4 | 2 | 1 | 4 | 19 | 9 |
| 5 | 4 | 7 | 11 | 7 | 6 |
| 6 | 2 | 8 | 15 | 7 | 3 |
| 7 | 1 | 4 | 18 | 10 | 2 |
| Total | 10 | 34 | 79 | 86 | 36 |
| Avg. | 4.08 | 13.87 | 32.24 | 35.1 | 14.6 |

Matching of interest of all (employer, market and employee) can be found and training (learning) can be started for leadership. 35.1% of the people were agreed to it and 14.6% strongly agreed to it.13.87% of the people disagreed and 4.08% strongly disagreed to it. While 32.24% remained neutral as shown in "fig. 3".

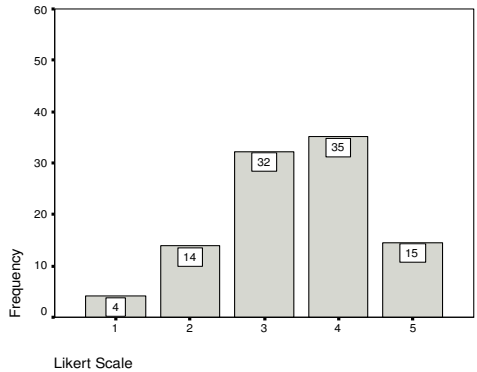

Figure 3. Graph showing the cumulative results of questionnaire for goal 3

*D. Cumulative Analysis of Goal 4. Effects of Proposed Solution.*

The learning for leadership results into enhancement of many leadership qualities such as technical skills, presentation, coordination and interaction. In addition to it this also provide the technical backup support. Thus making both organization and team members as market leaders. The cumulative result of survey of this goal is shown in table IV.

TABLE IV
CUMULATIVE ANALYSIS OF GOAL 4

| Q. No | Str. Disagree | Disagree | Neutral | Agree | Str. Agree |
|---|---|---|---|---|---|
| 1 |  | 1 | 4 | 16 | 14 |
| 2 |  | 2 | 5 | 20 | 8 |
| 3 | 1 | 1 | 2 | 20 | 11 |
| 4 |  |  | 7 | 19 | 9 |
| 5 |  | 4 | 4 | 6 | 21 |
| 6 |  | 1 | 9 | 10 | 15 |
| 7 |  | 1 | 4 | 14 | 16 |
| 8 | 1 | 3 | 2 | 17 | 12 |
| 9 |  | 2 | 8 | 19 | 6 |
| 10 | 9 | 18 |  | 4 | 4 |
| 11 | 6 | 9 | 7 | 6 | 7 |
| 12 | 3 | 14 | 11 | 3 | 4 |
| Total | 20 | 56 | 63 | 154 | 127 |
| Avg. | 4.76 | 13.33 | 15 | 36.66 | 30.23 |

Leadership qualities enhanced as a result of the learning for leadership. 36.66% of the people were agreed to it and 30.23% strongly agreed to it.13.33% of the people disagreed and 4.76% strongly disagreed to it. While 15% remained neutral as shown in "fig. 4".

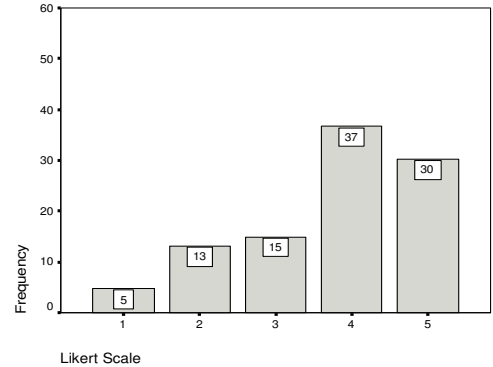

Figure 4. Graph showing the cumulative results of questionnaire for goal 4

VII. CONCLUSION

Software organizations must keep pace with the changing world of computers so they able to remain on the top of market and can sustain their success on long term basis. To be able to act as leader, continuous focus and refinement of leadership qualities is required. We focus not only the technical aspects of the leadership but also the other aspects such as coordination skills presentation skills, communication skills, motivation and satisfaction. Agile, match with our methodology, in the sense that both focuses on the people. Agile's focus on the people make it a natural choice to be used in leader organization that want to sustain this status in long run. Long term learning requirements of the organization can be found from the already in used documents I-e maintenance plan, marketing plan, maintenance contracts. In addition to it new emerging technologies must also explored for this purpose. We do not introduce any new document for this purpose because this will lead to increase time and cost of the software development lifecycle.

Further more agile value the running software over the comprehensive documentation [11] therefore another documentation was not introduced. We found following factors on the basis of above mentioned documents.
1. Future Projects
2. Domain of future projects
3. Time of maturity of projects
4. Learning requirements

All these factors are help full to predict the long term learning requirements of the future project. In this way they also set the line of action of the organization.

After finding the long term learning requirements of an organization for leadership, next step is to find the learning desires of the team members. Because if a team member is interested in .Net technologies and he/she is selected for java programming training he will not learn it whole heartedly.

Aptitude of learning toward specific learning desire can be judged during the daily Scrum meetings. To reduce the error team members can also be asked formally. This will give a clear picture of the learning desires of specific team members. Then learning opportunities are searched with in organization on the basis of long term learning requirements.

Now we have three lists which we use for finding persons to be trained; long term learning requirement of the organization for leadership, specific learning desires of the team members and learning opportunities with in that specific organization.

Learning desire list is matched with the long term learning requirements list and resulting intersection is further matched with the learning opportunities list to find a final list that is showing; who to train, where to train and why to train. The whole process is also depicted as

Learning for leadership= ((Learning requirements for leader ship ∩Learning desires) ∩ (learning opportunities))

The beauty of this approach is that it not only results into enhancement of the technical skills but also other leadership qualities. As team members move from team to team during learning for leadership process interaction among different teams increases. As team members are trained for future they remain satisfied and motivation level of them also increases.

During training in different teams members are presenting their ideas for example in scrum review meeting. This results into polishing of their presentation skills. Whole organization becomes dynamic and a culture of communication and coordination flourishes.

As people are trained in different teams knowledge keeps on flowing this results seamlessly in creation of a pool of the experts. These experts provide the back up support to the organization in case a person leaves, as team members are already familiar with technologies, culture and projects minimum efforts is required for replacement.

This work is supported with validation in which 63.19% people supported it while 16.46% disagreed to it as shown in "Fig. 5".

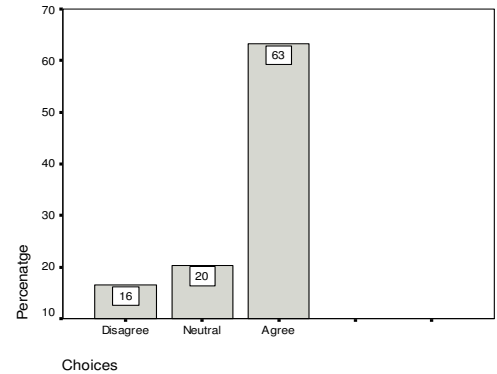

Figure 5. Graph showing the overall results of validation of proposed solution.